\documentclass[twocolumn,sn-mathphys-num]{sn-jnl}
\usepackage{geometry}
\usepackage{graphicx}
\usepackage{amsmath}
\usepackage{amssymb}

\usepackage[most]{tcolorbox}
\usepackage{listings}
\usepackage{booktabs}
\usepackage{multirow}
\usepackage{subcaption} % 必须引入此宏包以支持子表
\usepackage[version=4]{mhchem}
\usepackage{makecell}
\usepackage{siunitx}
\DeclareSIUnit\angstrom{\text {Å}}

\usepackage{bm}
\usepackage{url}

\usepackage[switch]{lineno}

\begin{document}

\title{Information Extraction of Nested Complex Structure of Quantum
Cascade Lasers via Large Language Models}

% A Novel JSON-Schema-Guided Prompting Strategy for Zero-Shot Extraction of Complex Hierarchical Data from scientific text

% A Novel JSON-Schema Guided Prompting Strategy for Zero-Shot Extraction of Complex Hierarchical Scientific Data: A Case Study on Quantum Cascade Lasers

% JSON-Schema Guided Information Extraction for Deeply Nested Nanostructures in Quantum Cascade Lasers

% Simple yet Robust: JSON-Schema Guided Information Extraction for Automated Quantum Cascade Laser Database Construction

% Beyond Traditional Prompting: A Schema-Constrained Pipeline for High-Fidelity Extraction of Nested Device Architectures

\author[1,3]{\fnm{Xiao} \sur{Fang}}

\author*[1]{\fnm{Ming} \sur{L\"u}}\email{mingl24@nudt.edu.cn}

\author[1]{\fnm{Hanwen} \sur{Liang}}

\author[1]{\fnm{Xingshen} \sur{Song}}

\author[2]{\fnm{Kele} \sur{Xu}}

\author[3]{\fnm{Hui} \sur{Cai}}

\author*[1]{\fnm{Chaofan} \sur{Zhang}}\email{c.zhang@nudt.edu.cn}

\affil[1]{\orgdiv{College of Advanced Interdisciplinary Studies}, \orgname{National University of Defense Technology}, \orgaddress{\city{Changsha}, \postcode{410073}, \country{China}}}

\affil[2]{\orgdiv{College of Computer Science and Technology}, \orgname{National University of Defense Technology}, \orgaddress{\city{Changsha}, \postcode{410073}, \country{China}}}

\affil[3]{\orgdiv{College of Computer}, \orgname{Nanjing University of Posts and Telecommunications}, \orgaddress{\city{Nanjing}, \postcode{210028}, \country{China}}}

\abstract{The rapid advancement of Large Language Models has transformed scientific research workflows,
including enabling the automated extraction of data directly from published literature.
Most existing efforts, however, focus on extracting simple labeled key-value entities,
whereas many scientific applications require more complex, hierarchically structured data.
A representative example is Quantum Cascade Lasers,
whose device architectures are defined by tens of interdependent parameters
organized in nested layer sequences.
In this work we propose a \emph{JSON-Schema Guided Information Extraction Pipeline}
(JSG-IE) that enables reliable extraction of deeply structured device data without model fine-tuning.
By transforming extraction into a schema-constrained generation task,
our approach significantly improves structural consistency and accuracy.
Across 12 state-of-the-art LLMs,
a properly designed JSON Schema improves performance by 5.7\% over conventional prompting,
with the highest $F_1$ score up to 83.4\%, achieved by the reasoning-enabled Kimi-k2-thinking model.
Importantly, this performance enhancement is most significant for mid-tier and open-source models,
where $F_1$ gains reach as high as 24.1\%,
effectively enabling these widely accessible models to achieve extraction fidelity previously restricted to much larger architectures.
This framework provides a scalable path toward automated construction of
high-fidelity device databases, accelerating data-driven optoelectronic design.}

% Our methodology and collected datasets have been made publicly available for the scientific community.


\maketitle
% \setboolean{displaycopyright}{false} % Do not include copyright or licensing information in submission.

% \linenumbers

\section{Introduction}
The emergence of AI for Science (AI4S) has initiated a profound paradigm shift in scientific discovery,
evolving from traditional empirical methods toward a closed-loop, data-driven architecture\cite{van2023ai,miret2025enabling,shao2026sciscigpt}.
Within this new landscape, Large Language Models (LLMs) act as pivotal cognitive engines,
distilling structured knowledge from vast, unstructured scientific literature.
Such a capability is particularly vital for the inverse design of complex optoelectronic devices,
where the scarcity of high-fidelity, large-scale training datasets remains a primary bottleneck.
A prominent class of such devices is the Quantum Cascade Laser (QCL)\cite{faist1994quantum},
one of the most widely adopted semiconductor light sources for the mid-infrared (MIR) to long-wave infrared (LWIR) spectral range\cite{beck2002continuous, gmachl2002ultra, bai2011room, hugi2012mid}.
A QCL typically comprises hundreds of epitaxially grown, nanometer-scale layers
engineered to form tailored electronic band structures that enable cascading
intersubband transitions and efficient carrier transport.
The thickness and material composition of each individual layer critically determine
the device’s emission wavelength, operating field, efficiency, and thermal performance.
Traditionally, QCLs structures are designed based on empirical design paradigms—such as
bound-to-bound\cite{faist1994quantum, faist1995continuous, sirtori2002mid, page1999high},
bound-to-continuum\cite{faist2001quantum, gmachl2002ultra, maulini2004broadband, lee2009broadband},
two-phonon resonance\cite{hofstetter2001high, beck2002continuous, blaser2005room, wittmann2009distributed},
and strong-coupling schemes\cite{liu2010highly}, et al.
% ✅此处需要一些参考文献，尤其是不同的设计思路的
% ✅考虑以 https://www.nature.com/articles/nphoton.2012.143 此和王俊普的博士论文为线索

Although notable efforts have been devoted to automated QCL design---ranging from conventional optimization algorithms\cite{franckie2020bayesian,bismuto2012fully}
to more recent machine learning (ML) techniques\cite{hernandez2022generating,hernandez2023application,correa2024machine,van2023ai,hu2023active,hu2024large}---both
paradigms encounter fundamental obstacles.
Optimization-based approaches are hindered by the highly nonlinear and non-differentiable nature of the transport and band-structure models of QCLs\cite{bismuto2012fully,mirvcetic2005towards},
whereas ML methods face the critical challenge of requiring large,
high-quality device databases---resources that remain scarce\cite{hernandez2022generating}.
% ✅ 看一下 https://ieeexplore.ieee.org/abstract/document/10089756 的参考文献列表里面有没有可以直接用的
While generating new experimental QCLs data is costly and time-consuming,
thousands of device structures have already been reported in the literature over the past three decades.
Unfortunately, these data are dispersed across prose, tables, and figures,
making systematic aggregation difficult.
As a result, leveraging prior design experience in a quantitative and scalable manner remains challenging.
Recent advances in Large Language Models (LLMs) offer a potential solution
by enabling automated extraction of structured information directly from scientific texts,
opening a path toward constructing comprehensive QCLs design databases.

Driven by this potential, the application of Information Extraction (IE) methods to
gather structured data from scientific literature has gained considerable momentum in recent years,
especially in materials and medical science\cite{wang2025nmrextractor, schilling2025text, dagdelen2024structured, wiest2024llm, chen2025large, bhattacharyya2025information, chen2025intelligent, sundaram2026automated}.
Historically, research in this domain has centered on core Natural Language Processing (NLP) tasks,
such as Named Entity Recognition (NER)\cite{goyal2025named, sidorova2025approach, zhang2025efficient},
Relation Extraction (RE)\cite{soltani2025llm}, and Knowledge Graph (KG) construction\cite{ateia2025llm}.
These methodologies are primarily designed to identify discrete entities and map pairwise relationships
within unstructured text to build domain-specific datasets.
The rapid advancement of LLMs,
including GPT-5\cite{sanli2025advances}, Claude\cite{anderson2025comparative}, and DeepSeek\cite{liu2025deepseek},
has further revolutionized the field by demonstrating exceptional natural language understanding and zero-shot extraction capabilities.
To balance performance with efficiency,
researchers have increasingly adopted Parameter-Efficient Fine-Tuning (PEFT \cite{peft}) techniques.
For instance, Zhang et al. developed a unified framework (LLM-UIE)
for low-resource domain adaptation\cite{zhang2025efficient},
while Song et al. utilized Low-Rank Adaptation (LoRA\cite{hu2022lora})
to significantly enhance NER performance \cite{song2025lora}.
However, these fine-tuning approaches impose substantial computational overhead and, more critically,
struggle to preserve strict structural integrity when extracting data with complex nesting and interdependent constraints--—such as
the layer sequences of QCLs, where parallel lists of materials, widths, and doping concentrations must remain precisely aligned.
Furthermore, the rapid pace of foundation model updates quickly renders fine-tuned domain models obsolete, eroding their practical value;
adapting PEFT methods to modern Mixture-of-Experts architectures and
their specialized sparse attention patterns also introduces nontrivial engineering challenges.
Consequently, although existing methods are effective for extracting isolated properties or
simple relational triples, they remain inadequate for the hierarchical and constraint-heavy extraction demands posed by QCLs.

% ✅这一段的主要内容应该是简单介绍如 QCL 的结构之间有如何相对复杂的嵌套结构，讨论此前 IE 针对简单标签的研究以及在这一问题上不适用，最终导出我们需要一个可推广的方法 (generalizable method) 来描述这一类问题并构建 prompt，最后说我们使用怎么样的 JSON schema with field-level iThesections to achieve this goal, 论证这种思路的优越性
The design of photonic devices like QCLs necessitates the reconstruction of nested,
hierarchical architectures in which layer sequences,
material compositions, and doping profiles are interdependent rather than isolated.
However, the above IE methodologies fail to meet these requirements in two key dimensions.
First, traditional NLP pipelines focus on planar entity-relation extraction,
which is limited to rigid data structures and cannot accommodate the multi-level nesting inherent in QCLs.
Second, while LLM-based IE offers a potential solution,
it typically suffers from limited cross-domain generalizability and high computational costs.
To bridge these gaps, a new prompting strategy is needed that shall describe
the logical and syntax requirement of the extracted data efficiently and effectively
both for an LLM to process the original literature files and for human researchers to
compose and edit.
Such approach shall transform the extraction task from an unconstrained generation problem
into a rigorous constraint-satisfaction problem,
achieving high-precision extraction of complex, nested data without resource-intensive model fine-tuning.

In this study, we introduce an approach for complex structured information extraction:
the \textbf{JSON-Schema Guided Complex Information Extraction Pipeline (JSG-IE)}.
This pipeline integrates JSON Schema with prompt engineering to simultaneously define
the target data structure and guide LLMs in extracting the structural information of QCLs.
The method allows researchers to flexibly define extraction templates for complex
nested parameters without the need for large-scale fine-tuning.
We systematically evaluate 12 state-of-the-art (SOTA) LLMs,
comparing different document preprocessing strategies and JSON structural templates.
Our results reveal that a properly designed JSON Schema, combined with reasoning-enabled models,
significantly enhances the accuracy and semantic consistency of the extracted data.
Moreover, this pipeline is model-agnostic and can be deployed vian LLM APIs,
enabling researchers to perform complex extraction tasks without deep expertise in LLM architecture.
By simply adjusting the JSON Schema structure and the \texttt{description} fields,
the system can be accurately adapted to a wide variety of complex structural extraction tasks.
Furthermore, the method ensures robust scalability,
enabling seamless adaptation to other specialized scientific domains with minimal reconfiguration.

\section{METHODS}

\begin{table*}[t]
\centering
% 强制压缩列间距
\setlength{\tabcolsep}{2pt}
\caption{Text extraction accuracy comparison of different PDF parsers.}
\label{tab:parsers}
\begin{tabular*}{\textwidth}{@{\extracolsep\fill}lcccccc}
\toprule
\textbf{Parser} &
\makecell[c]{\textbf{Cosine} \\ \textbf{Similarity}} &
\makecell[c]{\textbf{Jaccard} \\ \textbf{Similarity}} &
\makecell[c]{\textbf{Edit} \\ \textbf{Distance}} &
\makecell[c]{\textbf{2-gram}} &
\makecell[c]{\textbf{3-gram}} &
\makecell[c]{\textbf{Overall} \\ \textbf{Accuracy}} \\
\midrule
\textbf{OCRmyPDF} & \textbf{0.9978} & \textbf{0.9389} & \textbf{0.9788} & \textbf{0.9905} & \textbf{0.9601} & \textbf{0.9732} \\
PyPDF2 & 0.9899 & 0.8530 & 0.8819 & 0.9424 & 0.9072 & 0.9149 \\
pdfminer.six & 0.9850 & 0.8402 & 0.8011 & 0.9367 & 0.8908 & 0.8908 \\
pdfplumber & 0.9663 & 0.6138 & 0.2783 & 0.8516 & 0.7298 & 0.6880 \\
\botrule % 适配 sn-jnl 模板规范
\end{tabular*}
\end{table*}

\begin{table*}[htbp]
\centering
\caption{Performance Metrics Across Different Document Formats}
\label{tab:formats}
\begin{tabular}{lccc}
\toprule
\textbf{Format} & \textbf{Precision (\%)} & \textbf{Recall (\%)} & \textbf{$F_1$ Score (\%)} \\
\midrule
\textbf{Markdown} & \textbf{72.0} & \textbf{71.6} & \textbf{70.4} \\
PDF              & 65.5          & 64.6          & 65.0          \\
\LaTeX           & 47.0          & 45.0          & 45.0          \\
JSON             & 46.0          & 45.0          & 44.0          \\
\bottomrule
\end{tabular}
\end{table*}

\begin{figure*}[htbp]
    \centering
    \includegraphics[width=1\linewidth]{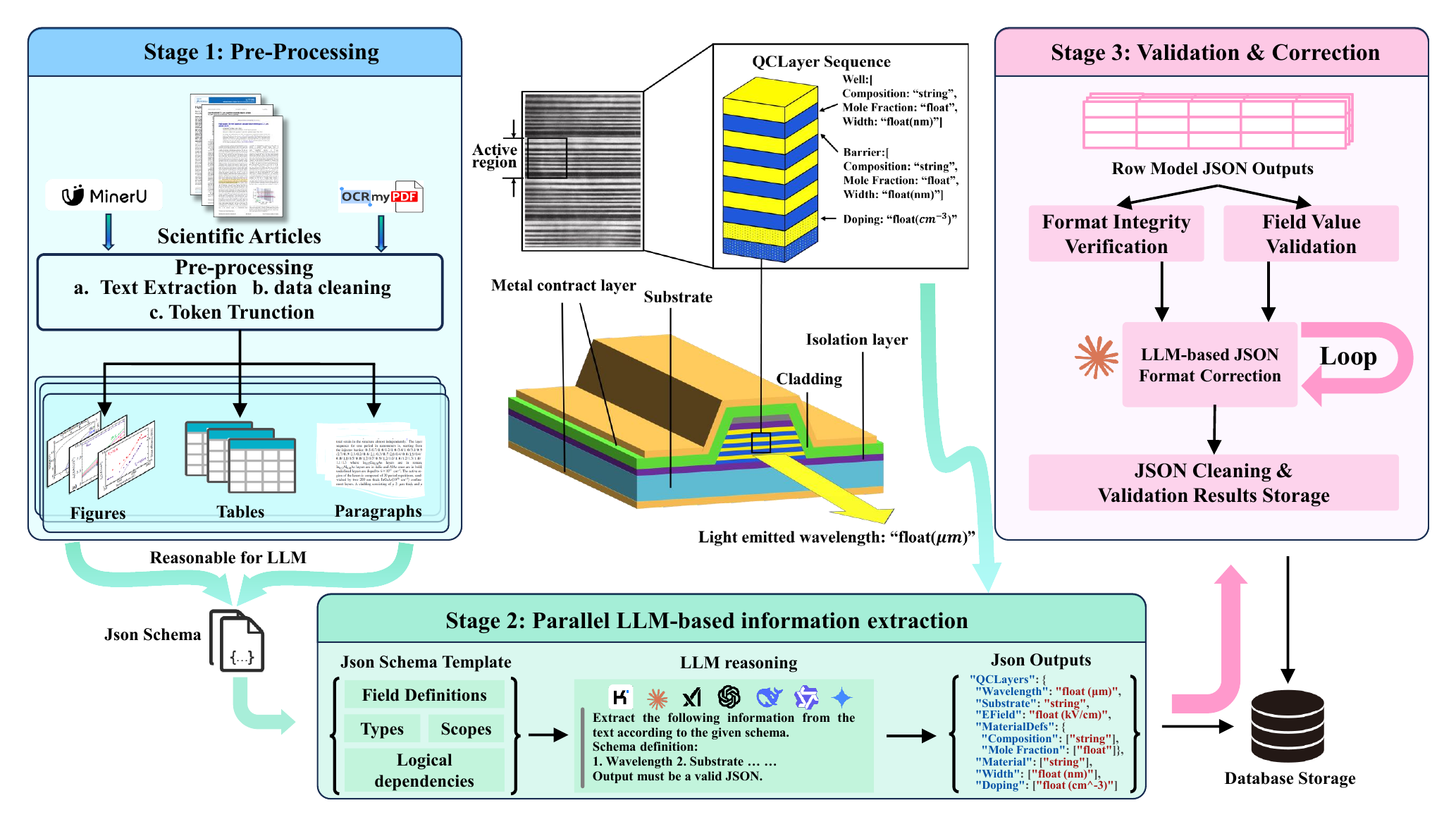}
    \caption{\textbf{Overview of the JSON-Schema Guided Information Extraction Pipeline.}
    The pipeline integrates three core stages for extracting information related to QCLs design.
    (1) Pre-processing: The pipeline initiates by processing relevant scientific articles
    on QCLs structural design using diverse PDF parsing methods.
    (2) Parallel LLM-based Information Extraction: Guided by JSON Schema,
    prompt engineering is employed to provide field-level instructions to the LLM for
    extracting parameters pertinent of QCLs.
    (3) Validation \& Correction: The raw LLM outputs undergo two-part validation,
    with errors corrected by LLM-based JSON format correction before final database storage.
    }
    \label{fig:pipelines}
\end{figure*}

\subsection{Data Curation}
The construction of a high-fidelity benchmark dataset is fundamental to evaluating the JSG-IE pipeline,
as QCL architectures are defined by dozens of interdependent parameters organized in deeply nested sequences.
We conducted a comprehensive literature search across major scientific databases,
including Web of Science and IEEE Xplore, covering the period from the inception of QCL technology in 1994 to early 2026.
While thousands of device structures have been reported over the past three decades,
these data remain dispersed across prose, tables, and figures,
making systematic aggregation exceptionally difficult.
To establish a reliable "ground truth," we implemented a rigorous filtering process that prioritized articles
providing a complete active region description—specifically requiring
emission wavelength, substrate material, and the full sequence of layer thicknesses and compositions.
However, to ensure absolute data integrity, each document was subjected to
a rigorous manual parsing and cross-verification process by researchers to verify structural consistency.
This intensive filtering process resulted in a final, high-quality benchmark dataset comprising 42 papers.

\subsection{Stage 1: Pre-processing}
As noted in the data curation process, the 42 curated scientific articles in our benchmark dataset are predominantly archived in PDF format.
This format presents significant challenges for LLMs due to its inherent lack of native structural support.
Our preliminary investigations into commercial OCR tools—intended to bridge this gap—revealed frequent layout distortions.
Specifically, in multi-column articles, some OCR processing may fail to recognize column boundaries,
resulting in an interleaved reading order where lines from adjacent columns are incorrectly merged.
Consequently, the primary objective of Stage 1 is to serialize these raw PDFs into a structured,
high-fidelity text representation that preserves the original functional hierarchy of the scientific information.

\subsubsection{Text Extraction}
The pipeline begins with the conversion of PDF content into plain text.
To identify the most reliable serialization method, we benchmarked several industry-standard PDF parsers,
evaluating their performance via text similarity metrics against the source documents.
As detailed in Table \ref{tab:parsers}, the OCRmyPDF package demonstrated superior reliability,
achieving a leading average accuracy of 97.32\%.
Consequently, OCRmyPDF was adopted for all subsequent PDF parsing tasks.

Furthermore, beyond simple text recovery, the specific representation of a document significantly influences the extraction performance of LLMs.
Utilizing the MinerU2.5 engine\cite{niu2025mineru25decoupledvisionlanguagemodel,wang2024mineruopensourcesolutionprecise,he2024opendatalab} , we evaluated the impact of various document formats, including PDF, LaTeX, JSON, and Markdown.
As shown in Table \ref{tab:formats}, Markdown emerged as the optimal format for this task, achieving an $F_1$ score of 70.4\%.
Markdown’s concise structure effectively minimizes syntactic redundancy—often referred to as "token noise"—inherent in LaTeX or JSON,
which tends to degrade model focus during complex information extraction.

\subsubsection{Priority-Based Truncation Strategy}
To accommodate the context window limitations of state-of-the-art LLMs,
we implemented a priority-based truncation strategy that ranks document sections by their functional relevance to the specific extraction task.
Under this framework, high-priority sections—specifically the Methods and Results sections detailing active region designs and layer sequences—are prioritized for retention.
Conversely, non-essential components such as references, prolonged introductions, and conclusions are systematically discarded.
This strategy ensures that the model’s finite attention is concentrated on the most informative segments of the literature,
mitigating information loss and reducing the risk of hallucination.

\subsection{Stage 2: Schema-Guided Information Extraction}
At the heart of the JSG-IE pipeline lies the proposed Schema-Guided Information Extraction framework.
This approach fundamentally transforms information extraction from a traditional task—often requiring continuous model fine-tuning into a natural language generation (NLG) problem.
By integrating JSON Schema with advanced prompt engineering,
the pipeline facilitates the reliable extraction of deeply nested device data from pre-processed text,
while bypassing the need for model training or specialized parameter updates.

\subsubsection{JSON Schema Design}

As illustrated in the central part of Fig.~\ref{fig:pipelines},
the core architecture of a QCL revolves around the design of the active region,
which consists of a periodical stack of coupled quantum wells and barriers.
The precise material composition and the spatial arrangement of these nanostructures are
the major parameters when designing a QCL active core structure,
as they dictate the electronic band and the intersubband transitions responsible for the light emission
and the electron transport.
However, capturing the parameters of these nanostructures from publications
are challenging because there is not a standard representation of the structure.
Conventional entity and relation extraction approaches are insufficient
for capturing the nested and hierarchical nature of such scientific information \cite{liu2025knowledge}.
Consequently, we designed the extraction pipeline to
transform unstructured text into structured JSON objects.
By integrating substrates and the active region sequences into a compound entity,
the pipeline ensures that the internal relationships and physical dependencies of the QCL device
are preserved as a meaningful, structured whole for a QCL database.

In the pipeline, we developed a comprehensive JSON Schema
that defines the semantics and constraints of QCLs data.
In our previous work \cite{lyu2021ErwinJr2, lyu2021software},
JSON is used as the standard serialization for the structure of a QCL device.
As detailed in Appendix~\ref{app:json_template},
JSON allows for the definition of customizable pre-extraction relationships
and accommodates the nested parameters inherent in complex device designs.
However, preliminary evaluations indicate that simple JSON templates often yield inaccurate results
when used directly as prompts. This is largely because raw JSON structures
can be semantically sparse within a long-context prompt,
making it difficult for LLMs to prioritize fine-grained details during extraction.
Drawing inspiration from constrained decoding research \cite{pezoa2016foundations},
we utilize JSON Schema as a rigorous framework for prompting.
By leveraging field-level instructions and structural constraints within a JSON Schema,
we transform the extraction task from an unconstrained generation problem into
a rigorous constraint-satisfaction problem.

The JSON schema not only describe the semantic meaning of the data for the QCL structure,
it may also implicitly regulate some logical constraint of the result.
For example, in QCLs, layer widths, the material and the composition of layers should be
lists of the same size.
We implemented two distinct structural formats to evaluate extraction efficiency:
a dictionary-of-list format (Fig.~\ref{fig:json_template_dict_of_list})
and a list-of-dictionary format (Fig.~\ref{fig:json_template_list_of_dict}).
The dictionary-of-list format is designed to consolidate QCLs layer attributes into
synchronized arrays, offering high machine-readability and direct compatibility
with the data structures used in most semiconductor simulation software.
However, this format poses a significant risk during LLM-based extraction:
it requires the model to maintain a strict one-to-one mapping across parallel lists,
where the omission of a single element can lead to index misalignment
and render the entire dataset unusable.
In contrast, the list-of-dictionary format encapsulates each layer
as a discrete object containing its specific attributes.
As shown in Table~\ref{tab:schema_comparison_2},
this format achieves a superior $F_1$ score (70.4\% vs. 65.2\%),
as it facilitates more precise, granular extraction.
Nonetheless, this preference is contextualized within complex IE tasks,
for data lacking extensive parallel mapping, dictionary-of-list remains a more streamlined alternative for serialization.

\subsubsection{Schema-Guided Prompt Engineering}

\begin{figure*}[htbp]
    \centering
    \includegraphics[width=1\linewidth]{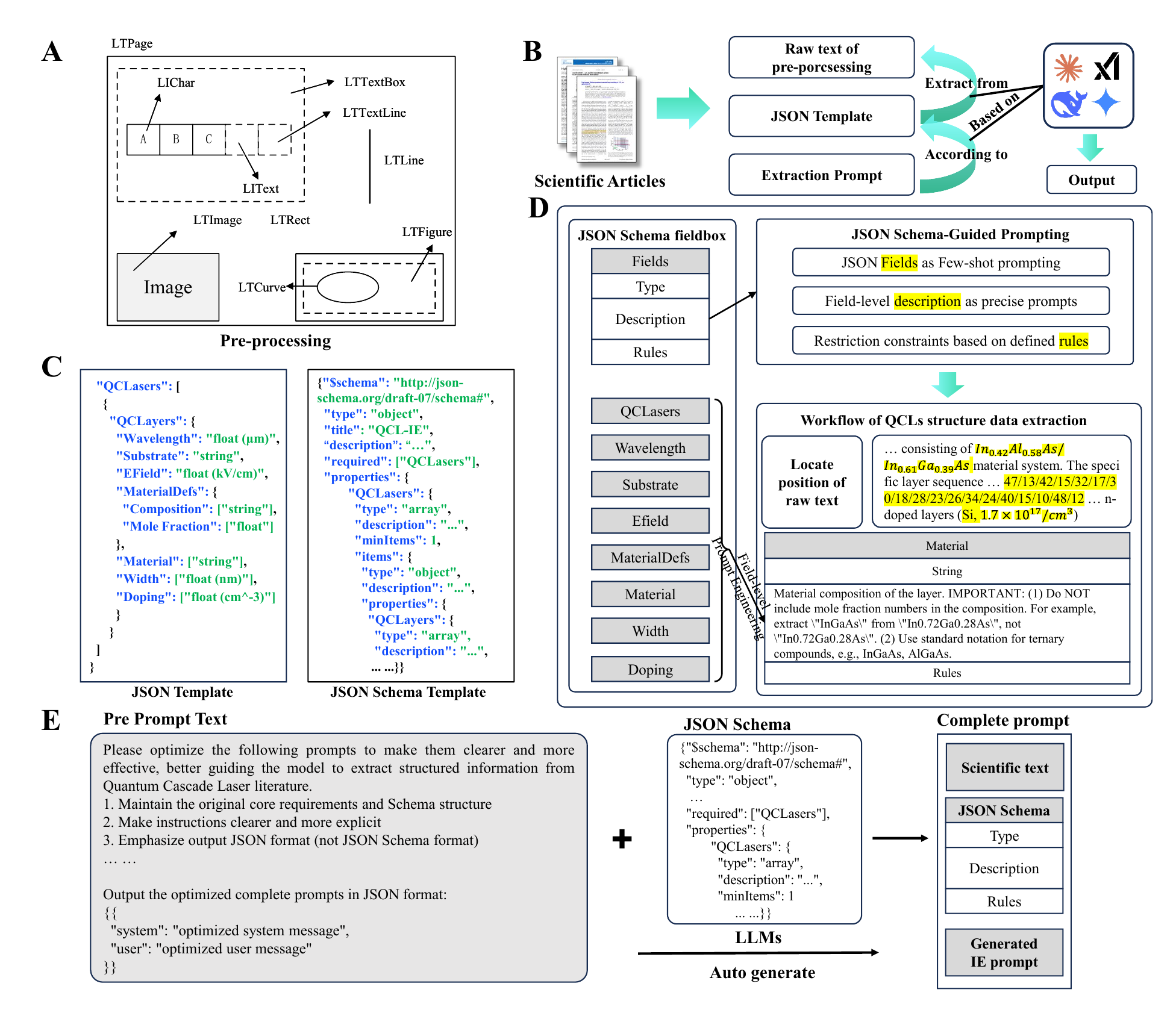}
    \caption{\textbf{JSON-Schema Guided Prompting} (A) Workflow of pre-processing for scientific articles. (B) Traditional prompt engineering for information extraction. (C) JSON and JSON Schema Template for QCLs Information Extraction. (D) Workflow of JSON-Schema Guided prompt engineering. (E) Complete prompt auto-generation}
    \label{fig:schema_prompting}
\end{figure*}

The core of our extraction strategy is to leverage the JSON Schema as the foundation for prompt engineering,
as illustrated in Fig.~\ref{fig:schema_prompting}.
Conventional prompt engineering for information extraction (Fig.~\ref{fig:schema_prompting}B)
typically integrates the raw text, a pre-defined JSON template, and instructional prompts
into a unified input for LLMs.
However, constraints within the context window inevitably result in information loss \cite{vuong2021does}.
Specifically, stringent requirements for specific field extractions may induce hallucinations,
thereby compromising extraction fidelity.
In contrast, as depicted in Fig.~\ref{fig:schema_prompting}D,
in our method, the data structure is defined by the JSON schema,
which includes names, types, and constraints, which provides a rigorous template for LLMs.
The \texttt{description} field provides granular instructions
that claim fine-grained semantic guidance for each field.
This approach transforms the extraction task from a simple "fill-in-the-blank" exercise
into a constraint-satisfaction problem while ensuring extraction accuracy.
A comparison of this schema-guided approach with conventional prompt engineering is presented in Table~\ref{tab:llm_performance_comparison}.
Fig.~\ref{fig:schema_prompting}E illustrates the workflow of the complete prompt generation process.
JSG-IE is designed to be the core of a fully automated information-extraction pipeline.
By specifying pre-prompt text and integrating the descriptive semantics defined in a JSON Schema,
the system uses LLMs to produce a complete, task-specific prompt for IE,
without manual drafting of extraction instructions.

% \subsection{API-based LLM Integration}
Prior studies suggest that fine-tuning LLMs on limited or noisy data can
compromise generalization capability \cite{hawkins2024effect, huang2024harmful, hahm2025unintended, huang2024booster}.
Given the potential for format inconsistencies and limited input sample in our extraction setting,
we opted against fine-tuning.
Instead, we accessed state-of-the-art LLMs via commercial APIs,
optimizing performance through systematic prompt design and parameter tuning.
Detailed model specifications and configuration parameters are listed in
Tables~\ref{tab:llm_apis} and \ref{tab:api_parameters}.

Furthermore, ensemble techniques, including consensus voting and stacking, is adopted
to integrate outputs from multiple LLMs.
However, as shown in Table~\ref{tab:llm_performance_comparison},
these ensemble mechanisms failed to surpass the performance of the best individual models,
consistently yielding results slightly below the best-performing single model.
Our analysis indicates that field-level independent voting disrupts
the inherent logic between correlated fields,
such as material types and their corresponding compositions.
This leads to a breakdown of the semantic consistency that a single model naturally maintains.
These findings suggest that conventional ensemble methods are insufficient for handling complex,
structured data.
Future works should explore sophisticated integration strategies that
preserve structural integrity and incorporate model-specific reliability awareness.

\begin{table*}[ht]
\centering
\caption{Technical specifications of the LLM APIs used in this study}
\label{tab:llm_apis}
\setlength{\tabcolsep}{4pt}
\begin{tabular}{@{}lccccc@{}}
\toprule
\textbf{Model} & \textbf{Parameters} & \textbf{Release} & \textbf{Context Window} & \textbf{Reasoning} \\
&  & \textbf{Date} & \textbf{(tokens)} & \textbf{Support} \\
\midrule
\textbf{claude-sonnet-4-5} & Undisclosed & 2025-09 & 200k / 1M & Yes \\
\textbf{deepseek-chat} & 671B & 2025-12 & 128K & No \\
\textbf{deepseek-v3-2} & 685B & 2026-01 & 164K & No \\
\textbf{deepseek-v3-2-thinking} & 685B & 2026-01 & 164K & Yes \\
% ✅ TODO: 2T+ 的意思是 ≥2T 吗？，如果是，这里改成 $\ge$2T 吧
\textbf{gemini-3-pro-preview} & $\ge$2T* & 2025-11 & 1M & Yes \\
\textbf{glm4-6} & 357B & 2025-09 & 256K & No \\
\textbf{gpt-4o-2024-11-20} & 1.8T* & 2024-11 & 128K & Yes \\
\textbf{gpt-5-chat} & $\ge$2.1T* & 2025-08 & 400K & Yes \\
\textbf{kimi-k2-0905-preview} & 1T & 2025-09 & 256K & Yes \\
\textbf{kimi-k2-thinking} & 1.2T & 2025-09 & 256K & Yes \\
\textbf{qwen3-max} & $\ge$1T* & 2025-09 & 256K & Yes \\
\textbf{qwen3-vl-235b-thinking} & 235B & 2025-12 & 32K & Yes \\
\bottomrule
\addlinespace
\multicolumn{5}{p{0.85\textwidth}}{\footnotesize \textit{Note: Parameters marked with an asterisk (*) are based on third-party statistics and industry estimates; official figures have not been disclosed by the developers.}}
\end{tabular}
\end{table*}

\begin{table*}[ht]
\centering
\caption{API inference parameters used for structured extraction}
\label{tab:api_parameters}
\begin{tabular}{@{}ll@{}}
\textbf{Parameter} & \textbf{Value} \\
\midrule
\textbf{temperature}       & 0.3    \\
\textbf{n}                 & 1      \\
\textbf{stream}            & false  \\
\textbf{top\_p}            & 0.95   \\
\textbf{max\_tokens}       & 4096   \\
\textbf{presence\_penalty}  & 0.5    \\
\textbf{frequency\_penalty} & 0.2    \\
\bottomrule
\end{tabular}
\end{table*}

\subsection{Stage 3: Validation \& Correction Loop}

% Post-processing is vital for robust information extraction, particularly regarding structured JSON outputs.
However, despite the structured guidance provided by JSON Schema,
we observed that the generated outputs still contained non-negligible errors in practice.
According to our experiments,
two primary categories of errors were identified in the generated JSON: format errors and value errors.
The former includes syntactic artifacts such as markdown code block tags (e.g., \texttt{```json}),
while the latter produces numerical data with incorrect decimal placement. For instance,
\texttt{"Width": [3.8, 13.0, …]} was incorrectly output as \texttt{"Width": [0.38, 1.3, …]},
which is likely due to incorrect model inference for unit \si{nm} and \si{\angstrom}.
To address these issues, Stage 3 employs a tiered correction strategy grounded in the same JSON Schema used for extraction.
Format errors are handled deterministically via a rule-based engine that
strips Markdown artifacts and verifies proper bracket enclosure without invoking the LLM.
For value errors, the engine first validates each field against the Schema’s numeric ranges and cross-field constraints
(e.g., equal array lengths for Material, Width, and Doping).
When violations are detected, the erroneous field—accompanied by diagnostic feedback specifying expected units,
ranges, or consistency rules—is re-submitted to the LLM for targeted re-extraction.
This closed-loop process ensures that both syntactic correctness and semantic validity are enforced systematically.

\section{RESULTS}

\subsection{Evaluation Criteria}
% ✅ TODO: 这节缺一个解释什么是baseline的内容
Extraction performance is evaluated using a hierarchical, field-level matching framework.
The schema attributes are represented as a set $\mathbb{F} = \{F_1, F_2, \dots, F_k\}$,
which includes top-level fields and nested sub-fields $f_i \in F_j$.
For a predicted field $F^\text{test}$ and the corresponding ground-truth $F^\text{true}$,
True Positives (TP) are defined as $F^\text{true} \cap F^\text{test}$, i.e.,
fields that are present and correctly matched in both.
False Positives (FP) and False Negatives (FN) are derived from set differences
$(F^\text{test} \setminus F^\text{true})$ and
$(F^\text{true} \setminus F^\text{test})$ respectively.
From these counts, precision, recall, and the $F_1$-score are computed as:

\begin{align}
&\text{Precision} (P) = \frac{\text{TP}}{\text{TP} + \text{FP}} \\
&\text{Recall} (R) = \frac{\text{TP}}{\text{TP} + \text{FN}} \\
&F_1 = 2 \cdot \frac{P \cdot R}{P + R}
\end{align}

% This set-theoretic logic applies recursively to complex objects like list-of-dictionary format (see Fig.~\ref{fig:json_template_list_of_dict}).

The matching criteria are data-type dependent:
numerical attributes (e.g., the bias electric field, the layer width and
the central gain wavelength) require exact numerical matching,
whereas string and array fields (e.g., the substrate material name and
the quantum well/barrier material name)
utilize normalized exact matching. For chemical formulations,
the evaluation treats different sequences
(e.g., \ce{In_{0.45}Al_{0.55}As} versus \ce{Al_{0.55}In_{0.45}As} as equivalent), provided the constituent elements and their corresponding mole fractions are identical.

The evaluation follows a hierarchical workflow.
Predicted QCL structures (each individual literature may include multiple QCL structures)
are first aligned with ground-truth counterparts based on
the gain wavelength and the structural similarity.
Within matched layers,
constituent fields like layer width and material type are scored independently.
Any unmatched layers are penalized as total FP or FN for all potential fields.
Finally, Precision, Recall,
and $F_1$-score are aggregated via macro-averaging across articles.
This approach assigns equal weight to each document, preventing articles with higher structural complexity from disproportionately biasing the system's performance metrics.

It is worth mentioning that, there is more than one way of defining the data structure
to represent the same QCL structure. The data type definition itself may encode
some extra information or inherent requirement. For example, in a QCL, the width
of each quantum layer can be represented by a list, so is the material composition
and doping density. These lists should be of the same size, if it is represented as
a ``dictionary of lists'' format.
Such logical constraint may be violated when the LLM fails to understand the inherent
relationship.
Another way of defining the structure is a "list-of-dictionary" pattern,
where each individual layer is represented as a dictionary with keys \texttt{Width},
\texttt{Material} and \texttt{Doping}.

To quantify the improvements offered by our proposed method,
we establish a baseline using a conventional zero-shot prompting strategy.
This baseline represents a standard extraction approach that does not utilize hierarchical schema guidance,
instead outputting data in the dictionary-of-list format.
By comparing our results against this baseline, we can isolate the performance gains specifically attributable to the JSG-IE pipeline.
In our experiments, to ensure the consistency of the evaluation,
the data represented as a dictionary of lists (including the baseline)
is converted into a list of dictionaries.
Subsequently, a uniform comparison is conducted using the list of dictionaries format.
The corresponding comparison results are presented in Table~\ref{tab:llm_performance_comparison}.
Furthermore, a comparison of the baseline data within the original dictionary of
lists scale is provided, with detailed information available in Table~\ref{tab:schema_comparison_2}.

\subsection{Performance}\label{subsec:performance_extraction}
Table~\ref{tab:llm_performance_comparison} and Table~\ref{tab:average_gains}
provide a comprehensive evaluation of 12 state of the art LLMs,
contrasting our proposed JSG-IE pipeline against traditional instruction-based prompting.
The performances are calculated with an extract word match basis,
where the ground-truth data are collected and verified by humans.
%% ✅ data 这个词严格来说是个复数名词，单数形式datum。在日常英语中 data are 已经很常见了，但学术上多数时候还是用 data are
The results in Table~\ref{tab:llm_performance_comparison} show that JSG-IE
consistently outperforms the baseline across nearly all metrics,
achieving a mean $F_1$ score of 70.4\% and a 5.7\% absolute improvement
over the baseline (64.7\%). In our experiments,
\textbf{kimi-k2-thinking} reached the highest $F_1$ score of 83.4\%
under the JSG-IE configuration.

\begin{table*}[t]
\centering
\caption{Performance Comparison between  and JSON Schema Guided Extraction}
\label{tab:llm_performance_comparison}
\small
\begin{tabular*}{\textwidth}{@{\extracolsep\fill}l | ccc | ccc | c}
\toprule
\multirow{2}{*}{\textbf{Model Name}} & \multicolumn{3}{c|}{\textbf{Baseline}} & \multicolumn{3}{c|}{\textbf{JSG-IE}} & \textbf{Avg} \\
\cmidrule(lr){2-4} \cmidrule(lr){5-7} \cmidrule(lr){8-8}
& \textbf{P (\%)} & \textbf{R(\%)} & \textbf{$F_1$(\%)} & \textbf{P(\%)} & \textbf{R(\%)} & \textbf{$F_1$(\%)} & $F_1$(\%)\\
\midrule
\textbf{integrated}               & \pmb{79.4} & 73.4 & \pmb{75.2} & \pmb{84.0} & 82.2 & 81.5 & \pmb{78.4} \\
\midrule
\textbf{kimi-k2-thinking}        & 76.7 & \pmb{74.2} & 73.0 & 82.2 & \pmb{87.5} & \pmb{83.4} & 78.2 \\
\textbf{gemini-3-pro-preview}    & 78.5 & 70.6 & 72.5 & 73.8 & 73.8 & 73.1 & 72.8 \\
\textbf{deepseek-v3-2-thinking}  & 72.8 & 70.6 & 69.4 & 81.1 & 72.3 & 71.9 & 70.7 \\
\textbf{deepseek-chat}           & 69.4 & 69.3 & 68.2 & 63.1 & 61.3 & 61.3 & 64.8 \\
\textbf{gpt-5-chat}              & 68.8 & 68.1 & 67.3 & 57.4 & 60.1 & 57.5 & 62.4 \\
\textbf{deepseek-v3-2}           & 70.3 & 67.4 & 66.9 & 70.8 & 70.6 & 69.5 & 68.2 \\
\textbf{claude-sonnet-4-5}       & 60.3 & 62.2 & 60.8 & 63.5 & 63.9 & 63.4 & 62.1 \\
\textbf{qwen3-vl-235b-thinking}  & 65.0 & 61.0 & 59.8 & 78.6 & 80.2 & 77.5 & 68.7 \\
\textbf{qwen3-max}               & 58.8 & 60.1 & 58.7 & 65.9 & 69.5 & 65.8 & 62.3 \\
\textbf{kimi-k2-0905-preview}    & 65.3 & 50.2 & 54.2 & 75.2 & 71.6 & 72.6 & 63.4 \\
\textbf{gpt-4o-2024-11-20}       & 57.3 & 53.1 & 53.9 & 61.4 & 63.1 & 62.1 & 58.0 \\
\textbf{glm4-6}                  & 55.1 & 49.9 & 51.7 & 78.6 & 74.9 & 75.8 & 63.8 \\
\midrule
\textbf{Average}                 & \textbf{68.0} & \textbf{63.9} & \textbf{64.7} & \textbf{72.0} & \textbf{71.6} & \textbf{70.4} & \textbf{67.6} \\
\bottomrule
\end{tabular*}
\end{table*}

\begin{table*}[htbp]
\centering
\caption{Average Performance Gains Across Different Dimensions}
\label{tab:average_gains}
\begin{tabular}{l | ccc}
\toprule
\textbf{Comparison Dimension} & \textbf{P (\%)} & \textbf{R (\%)} & \textbf{$F_1$ (\%)} \\
\midrule
JSG-IE vs. Base$^1$    & +4.0 & +7.7 & +5.7 \\
List vs. Dict$^2$      & +1.9 & +3.1 & +1.8 \\
Markdown vs. PDF$^3$  & +4.6 & +6.8 & +5.7 \\
\bottomrule
\end{tabular}
\vspace{0.1cm}
\begin{flushleft}
\footnotesize
% \centering
$^1$ Difference between JSG-IE and Baseline averages in Markdown.\\
$^2$ Difference between List-of-Dictionary and Dictionary-of-List averages.\\
$^3$ Comparison between Markdown and PDF modalities under JSG-IE.
\end{flushleft}
\end{table*}

Integrating multiple LLMs improves the precision can be further improved
beyond that of the best individual model, with slight degradation in recall.
The majority-voting consensus strategy \cite{meyen2021group} is adopted % ✅TODO: cite for majority-voting consensus
to integrate multi-model result, labeled as the "integrated" column in these tables.
The integrated result achieves the highest precision in both the baseline (79.4\%) and JSG-IE (84.0\%).
That is because the voting logic prioritizes fields with the highest frequency of occurrence,
leading to a substantial increase of the precision $P$, which is a result of reduction in FP,
while TP remain relatively stable.
However, the recall $R$ remains limited or slightly decreases,
because correct answers uniquely identified
by a minority of models may be overruled by a majority of incorrect votes.
While the integrated method ensures superior output reliability,
its ability to maximize total information coverage is somewhat diluted
by this consensus-driven trade-off.

Although semantically equivalent, a schema that logically more compact
performs better in most LLMs.
Table~\ref{tab:average_gains} shows the efficacy of the two different schema design of JSG-IE,
the list-of-dictionary format and the dictionary-of-list format.
The list-of-dictionary format yields a 1.8\% higher $F_1$ score than
the dictionary-of-list format.
% ✅ TODO: "conducted on the dictionary-of-list scale" 想表达什么意思？
Furthermore, we extended our evaluation to include the original dictionary-of-list representation,
as detailed in Table~\ref{tab:schema_comparison_2}.
The results indicate that the list-of-dictionary format consistently
outperforms the dictionary-of-list alternative across nearly all tested models.
This performance gap likely stems from the attention mechanisms of contemporary LLMs;
the list-of-dictionary format aligns more closely with the logical grouping of
physical components in a QCL device,
enabling the model to focus on one discrete attribute set at a time
rather than managing multiple parallel lists.

Moreover, the significant performance gap can be observed for different input format.
%TODO: 这段是否考虑把 LaTeX等其他格式纳入讨论？
In our test, pre-processing PDF files to Markdown increases the $F_1$ score by 5.7\%
compared to raw PDF input,
underscores the critical role of document hierarchy.
This disparity primarily stems from information misalignment and
loss during the text extraction of PDF.
Specifically, our empirical observations indicate that multi-column layouts
in PDF are frequently misinterpreted as single-column text,
leading to severe disruptions in the reading order of the document.
In contrast, the hierarchical structure of Markdown effectively
preserves the core content of the research papers,
providing a cleaner contextual anchor that is more readily understood by the model.
Collectively, these results suggest that the synergy between schema-based guidance and
structural text input is pivotal for high-fidelity information extraction
across diverse LLM architectures, as detailed in Table~\ref{tab:performance_gains_analysis_detail}.

\begin{table*}[htbp]
\centering
\caption{Absolute Gains in $F_1$ Score Attributed to Thinking Capabilities}
\label{tab:thinking_gains_f1}
\begin{tabular}{@{}lccc@{}}
\toprule
\textbf{Model Group} & \textbf{Baseline} & \textbf{Dict of List} & \textbf{List of Dict} \\
\textbf{(Thinking vs. Standard)} & \textbf{($\Delta F_1$)} & \textbf{($\Delta F_1$)} & \textbf{($\Delta F_1$)} \\ \midrule
DeepSeek-v3-2 Group           & +2.5             & +6.0                 & +2.4                 \\
Kimi-k2 Group                 & +18.8            & +7.5                 & +10.8                \\
Qwen3 Group                   & +1.1             & +3.5                 & +11.7                \\ \midrule
\textbf{Average Thinking Gain} & \textbf{+7.5}    & \textbf{+5.7}        & \textbf{+8.3}        \\ \bottomrule
\end{tabular}
\end{table*}

Within the results from Table~\ref{tab:thinking_gains_f1},
we conclude that a substantial "reasoning premium" among models with enhanced thinking capabilities.
Reasonable models exhibit an average $F_1$ gain of 5.7\% to 8.3\%.
Critically, we observe that the baseline performance of deepseek-v3-2-thinking ($F_1$ of 69.4\%) surpasses that of
several standard models even when the latter are equipped with full JSG-IE guidance,
which suggests that internal reasoning chains can partially simulate external structural constraints,
allowing the model to maintain high extraction accuracy by
deducing logical relationships between scientific parameters,
such as matching the active region material of a QCL to its emission wavelength,
even in raw-instruction environments.

% \section{DISCUSSION}

\subsection{Analysis of Model Performance Heterogeneity}

\begin{table*}[t]
\centering
\caption{Individual $F_1$ score gains via the JSG-IE pipeline.}
\label{tab:individual_f1_gains}
\small
\begin{tabular*}{\textwidth}{@{\extracolsep\fill}lccc}
\toprule
\textbf{Model} & \textbf{Baseline $F_1$ (\%)} & \textbf{JSG-IE $F_1$ (\%)} & \textbf{$\Delta F_1$ (\%)} \\
\midrule
\textbf{Integrated} & 75.2 & 81.5 & +6.3 \\
\midrule
\textbf{Kimi-k2-thinking} & 73.0 & 83.4 & +10.4 \\
\textbf{Gemini-3-pro-preview} & 72.5 & 73.1 & +0.6 \\
\textbf{DeepSeek-v3-2-thinking} & 69.4 & 71.9 & +2.5 \\
\textbf{DeepSeek-chat} & 68.2 & 61.3 & -6.9 \\
\textbf{GPT-5-chat} & 67.3 & 57.5 & -9.8 \\
\textbf{DeepSeek-v3-2} & 66.9 & 69.5 & +2.6 \\
\textbf{Claude-sonnet-4-5} & 60.8 & 63.4 & +2.6 \\
\textbf{Qwen3-vl-235b-thinking} & 59.8 & 77.5 & +17.7 \\
\textbf{Qwen3-max} & 58.7 & 65.8 & +7.1 \\
\textbf{Kimi-k2-0905-preview} & 54.2 & 72.6 & +18.4 \\
\textbf{GPT-4o-2024-11-20} & 53.9 & 62.1 & +8.2 \\
\textbf{GLM4-6} & 51.7 & 75.8 & +24.1 \\
\bottomrule
\end{tabular*}
\end{table*}

While the JSG-IE pipeline yields a mean $F_1$ gain of 5.7\% across all 12 tested models, this aggregate metric obscures a more practical finding: the framework acts as a capability equalizer. Table~\ref{tab:individual_f1_gains} shows that the most pronounced improvements occur in mid-tier and open-source models that initially performed poorly under standard zero-shot prompting. For instance, GLM4-6 and Kimi-k2-0905-preview saw $F_1$ increases of 24.1\% and 18.4\% respectively, moving them from unreliable scores around 50\% to production-ready levels above 70\%. The JSON Schema appears to supply essential structural scaffolding that offsets the weaker internal reasoning chains in these models, allowing them to manage the high cognitive load of nested device architectures.

A different pattern emerged with top-tier architectures such as GPT-5-chat ($-9.8\%$) and DeepSeek-chat ($-6.9\%$), where performance declined. One possible explanation is that these advanced models already maintain robust internal reasoning chains that can partially simulate structural constraints. When burdened with exhaustive field-level descriptions and rigid schema constraints, the explicit instructions may introduce token-level noise that interferes with the model's natural inference path. From a practical standpoint, the value of JSG-IE lies less in boosting already-strong models than in enabling smaller, broadly accessible models to achieve high-fidelity extraction, which matters directly for building scalable and cost-effective scientific databases.

\section{DISSCUSSION}
The JSG-IE pipeline extracts complex QCL device structures from the literature without domain-specific fine-tuning, improving the average $F_1$ score by 5.7\% over standard prompt engineering. A closer look at the distribution of gains, however, reveals a more nuanced interaction between structural guidance and different LLM architectures.

\begin{itemize}
    \item The most notable effect is that JSG-IE functions as a capability equalizer rather than a universal booster. As Table~\ref{tab:individual_f1_gains} shows, mid‑tier models such as GLM4‑6 and Kimi‑k2‑0905‑preview gain 24.1\% and 18.4\% in $F_1$, moving from unreliable accuracy around 50\% to production‑ready levels above 70\%. The JSON Schema supplies the structural scaffolding these models lack for nested device layouts, compensating for their weaker internal reasoning chains. In contrast, some top‑tier Chain‑of‑Thought models, including GPT‑5‑chat, perform slightly worse under JSG‑IE. A plausible explanation is that these architectures already simulate structural constraints internally; the additional field‑level descriptions introduce token‑level noise that interferes with the model’s natural inference path. For practical database construction, this means high‑fidelity extraction can be achieved with broadly accessible, cost‑effective models, without depending exclusively on the most resource‑intensive ones.

    \item The superiority of the list‑of‑dictionary format (a 1.8\% higher $F_1$) underscores the value of aligning output structures with the localized attention mechanisms of LLMs. In QCL design, a layer’s thickness is meaningful only in combination with its material and doping. Parallel arrays, while compact, force the model to maintain strict one‑to‑one mapping across long, disconnected sequences, which makes index‑shift errors likely. By encapsulating all attributes of a single physical layer in one dictionary, the pipeline reduces cognitive load and allows the model to focus on one semantically coherent unit at a time.

    \item The 5.7\% $F_1$ advantage of Markdown over raw PDF input (Table~\ref{tab:average_gains}) confirms that document preprocessing sets the upper bound for extraction quality. Standard PDF parsers often mishandle multi‑column layouts, interleaving text in ways that break logical flow. Markdown’s hierarchical markers preserve a clean, sequential context that complements schema‑based generation. Together with the previous findings, this indicates that combining structured text inputs with explicit output schemas forms a robust, generalizable path for building automated databases in specialized scientific domains.
\end{itemize}

\section{CONCLUSION}
In this study, we presented the JSG-IE pipeline,
a robust framework designed to extract complex and hierarchical device structures of QCLs
from unstructured scientific literature.
By transforming the extraction task from an unconstrained generation problem
into a rigorous constraint-satisfaction problem,
JSG-IE eliminates the necessity for resource-intensive model fine-tuning.
Our systematic evaluation of 12 state-of-the-art LLMs demonstrates that
the integration of field-level instructions within a proper JSON Schema
significantly enhances extraction fidelity.
Under this framework, the reasoning-enabled \textbf{kimi-k2-thinking} achieved
a SOTA peak $F_1$ score of 83.4\%,
representing a substantial improvement over conventional instruction-based prompting.

Our findings underscore three pivotal elements for high-precision scientific information extraction:
high-fidelity document representation (Markdown), design of JSON Schema,
and the "reasoning premium" inherent in CoT enabled architectures.
Notably, our experimental results demonstrate that
advanced reasoning models can partially mitigate the absence of explicit structural guidance
through internal logical deduction.
While the majority-voting consensus mechanism achieved superior precision (84.0\%),
it failed to yield a corresponding improvement in the overall $F_1$ score,
primarily due to the conservative suppression of recall.
Future research may focus on developing sophisticated ensemble strategies,
such as confidence-aware integration or the strategic exclusion of underperforming models,
to construct a more robust and high-performance information extraction pipeline.

Our findings also reveal that while advanced reasoning models can partially mitigate the absence of explicit structural guidance through internal logical deduction, our pipeline allows mid-tier and open-source models to overcome their inherent reasoning limitations. Specifically, JSG-IE enabled models like GLM4-6 to gain 24.1\% in $F_1$. An interesting direction for future work would be to explore whether lightweight schema variants, or adaptive constraint injection, can retain the equalizing benefits for smaller models while avoiding the interference observed in stronger ones.

Overall, the JSG-IE pipeline offers a scalable and model-agnostic solution that
can be rapidly deployed via standard APIs and customized for diverse research domains.
By bridging the gap between dispersed literature and structured device databases,
this work provides a foundational tool for data-driven optoelectronic design.
Our implementation and evaluation datasets is made publicly available to
support the academic community in accelerating the automated synthesis of complex scientific knowledge.

% \newpage

\section{RESOURCE AVAILABILITY}
Requests for further information and resources should be directed to
and will be fulfilled by Ming L\"u (mingl24@nudt.edu.cn).

\subsection{Materials availability}

This study did not generate new materials.

\subsection{Data and code availability}

The research artifacts, including the JSON Schema template,
evaluation datasets (validation set),
and the core implementation code for prompt generation and extraction,
are available at \url{https://github.com/fangtoast/JSON-Schema-Guided-Information-Extraction-Pipeline}.
Due to copyright restrictions,
the full text of the original research papers used in this study is not provided.
Furthermore, API keys and private model configurations are withheld to
ensure security and compliance with service terms.
Additional experimental details will be made available on request.

\section{ACKNOWLEDGMENTS}
This work is all funded by Innovation Research Foundation of National University of Defense Technology.

\section*{AUTHOR CONTRIBUTIONS}
Xiao Fang: Writing – original draft, Software, Methodology, Data Curation;
Ming L\"u: Writing – review \& editing, Conceptualization, Methodology;
Hanwen Liang: Data Curation, Investigation;
Xingshen Song: Writing – review \& editing;
Kele Xu: Writing – review \& editing;
Hui Cai: Methodology;
Chaofan Zhang: Methodology, Project administration \& Funding.

\section{DECLARATION OF INTERESTS}
The authors declare that they have no known competing financial interests
or personal relationships that could have appeared to influence the work reported in this paper.

% \newpage

\bibliography{references}

\newpage

\appendix
\section{JSON Schema Design}\label{app:json_schema_dict_of_list}

\subsection{JSON Template for QCLs Information Extraction}\label{app:json_template}
\begin{figure}[!htbp]
  \centering
  \includegraphics[width=0.75\linewidth]{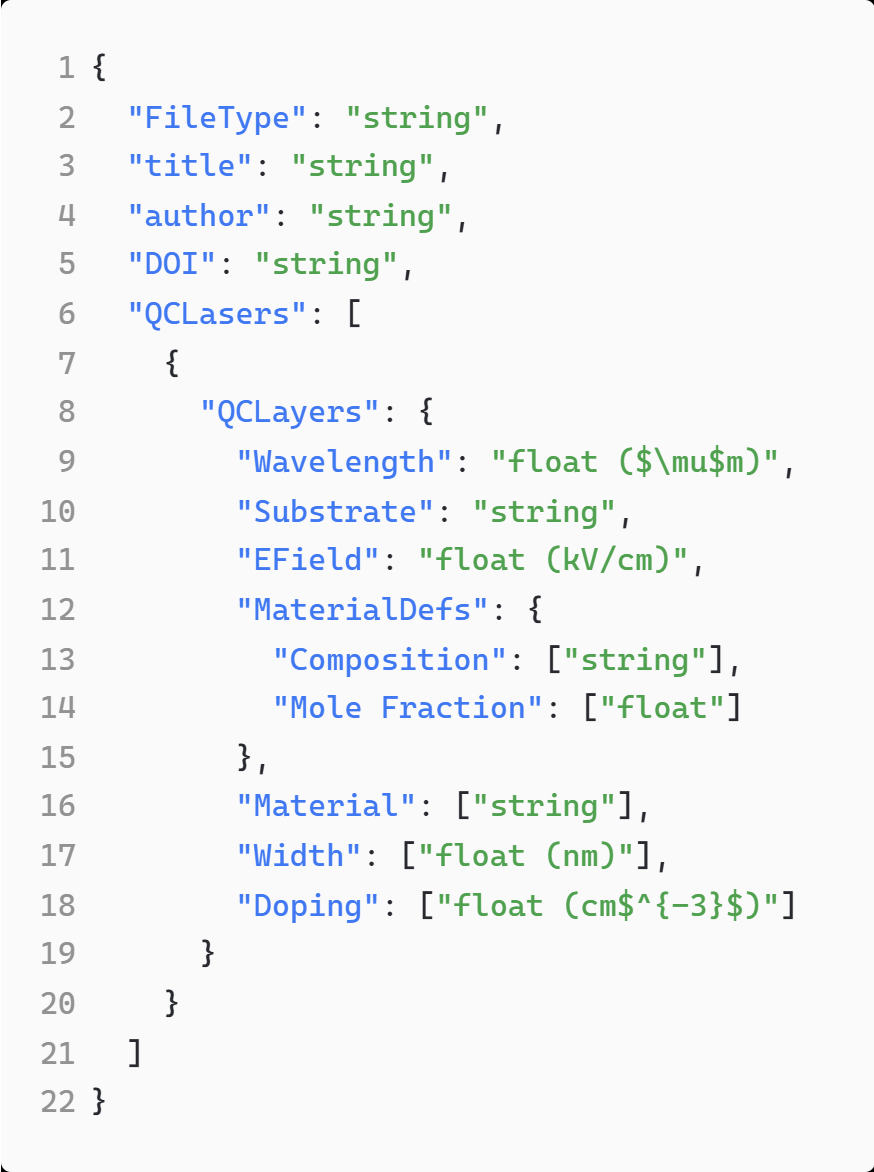}
  \caption{
  JSON template for structured extraction of QCLs device parameters.
  The template defines the hierarchical structure for
  capturing document metadata and device information including material compositions,
  layer dimensions, and doping profiles.}
  \label{fig:json_template_dict_of_list}
\end{figure}

Fig.~\ref{fig:json_template_dict_of_list} illustrates the designed JSON template for
extracting Quantum Cascade Laser (QCLs) structural information from scientific literature.
This template is hierarchically organized to capture both document-level metadata
and detailed device-specific parameters.

\begin{figure}[!htbp]
  \centering
  \includegraphics[width=0.75\linewidth]{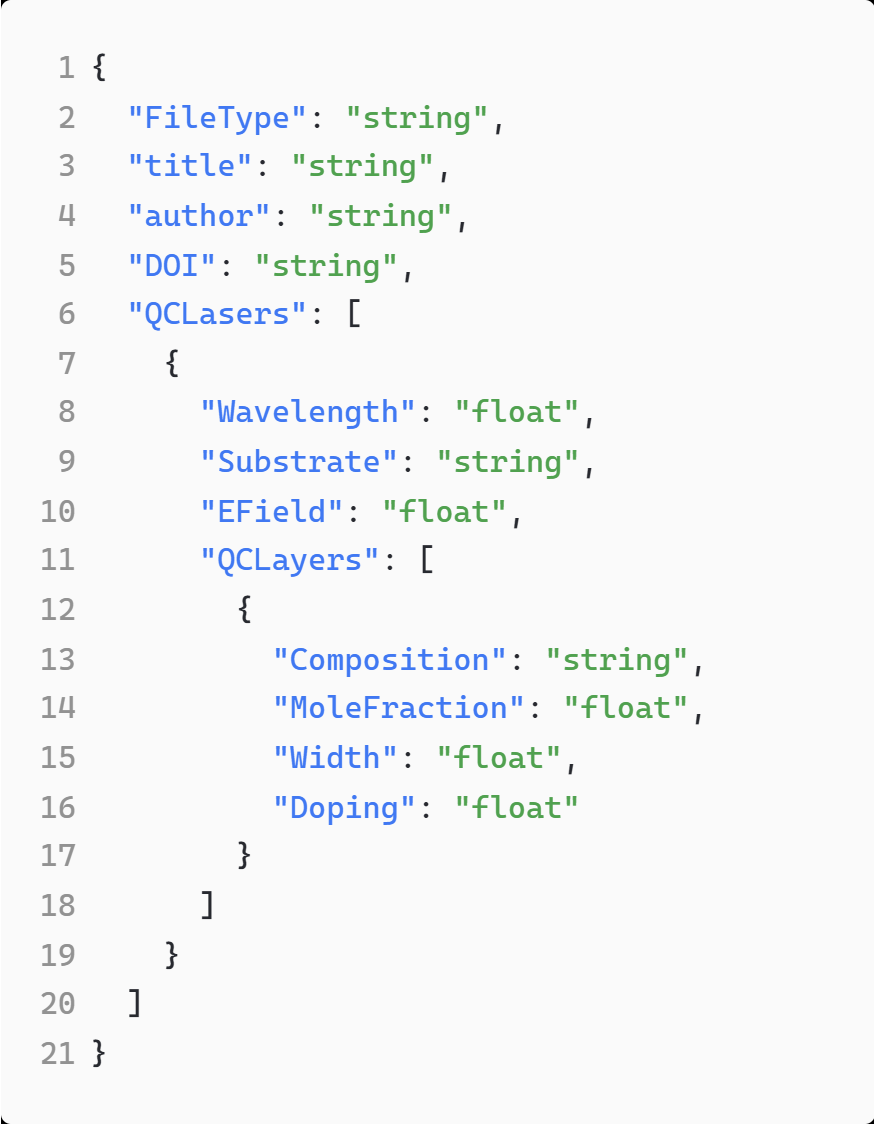}
  \caption{
  JSON template for structured extraction of QCLs device parameters.
  The template defines the hierarchical structure for capturing document metadata and
  device information including material compositions, layer dimensions, and doping profiles.}
  \label{fig:json_template_list_of_dict}
\end{figure}

\subsection{Field Descriptions}

Each field in the JSON template serves a specific purpose in capturing QCLs-related information:

\begin{itemize}
    \item \textbf{FileType}: Document format identifier (e.g., ``journal-article'', ``conference-paper'')
    \item \textbf{title}: Article title
    \item \textbf{author}: Author list
    \item \textbf{DOI}: Digital Object Identifier
    \item \textbf{QCLasers}: Array containing one or more QCLs device descriptions
    \begin{itemize}
        \item \textbf{QCLsayers}: Core device parameters
        \begin{itemize}
            \item \textbf{Wavelength}: Emission wavelength in micrometers ($\mu$m)
            \item \textbf{Substrate}: Growth substrate material
            \item \textbf{EField}: Operating electric field in kV/cm
            \item \textbf{MaterialDefs}: Alloy composition definitions
            \begin{itemize}
                \item \textbf{Composition}: Chemical formulas (e.g., "\ce{AlGaAs}", "\ce{InGaAs}")
                \item \textbf{Mole Fraction}: Corresponding stoichiometric fractions
            \end{itemize}
            \item \textbf{Material}: Array of materials in layer stack
            \item \textbf{Width}: Layer thicknesses in nanometers (nm)
            \item \textbf{Doping}: Impurity concentrations in cm$^{-3}$
        \end{itemize}
    \end{itemize}
\end{itemize}

\subsection{Complete JSON Schema Definition}

\begin{figure*}[t]
  \centering
  \includegraphics[width=\linewidth]{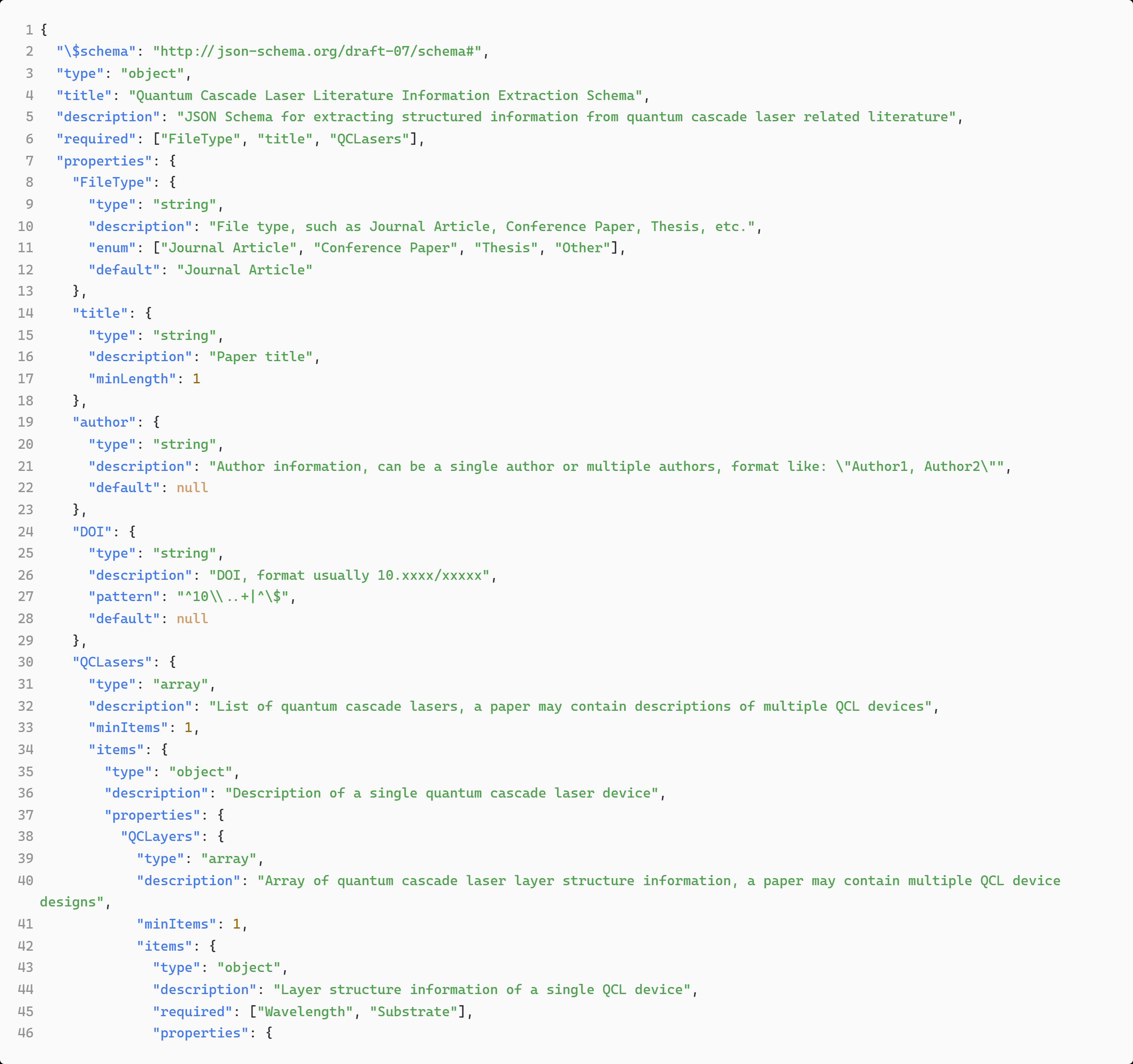}
  \caption{
  JSON Schema definition for QCLs literature information extraction (Part I):
  Top-level structure and metadata.}
  \label{fig:json-schema-part1}
\end{figure*}

\begin{figure*}[t]
  \centering
  \includegraphics[width=\linewidth]{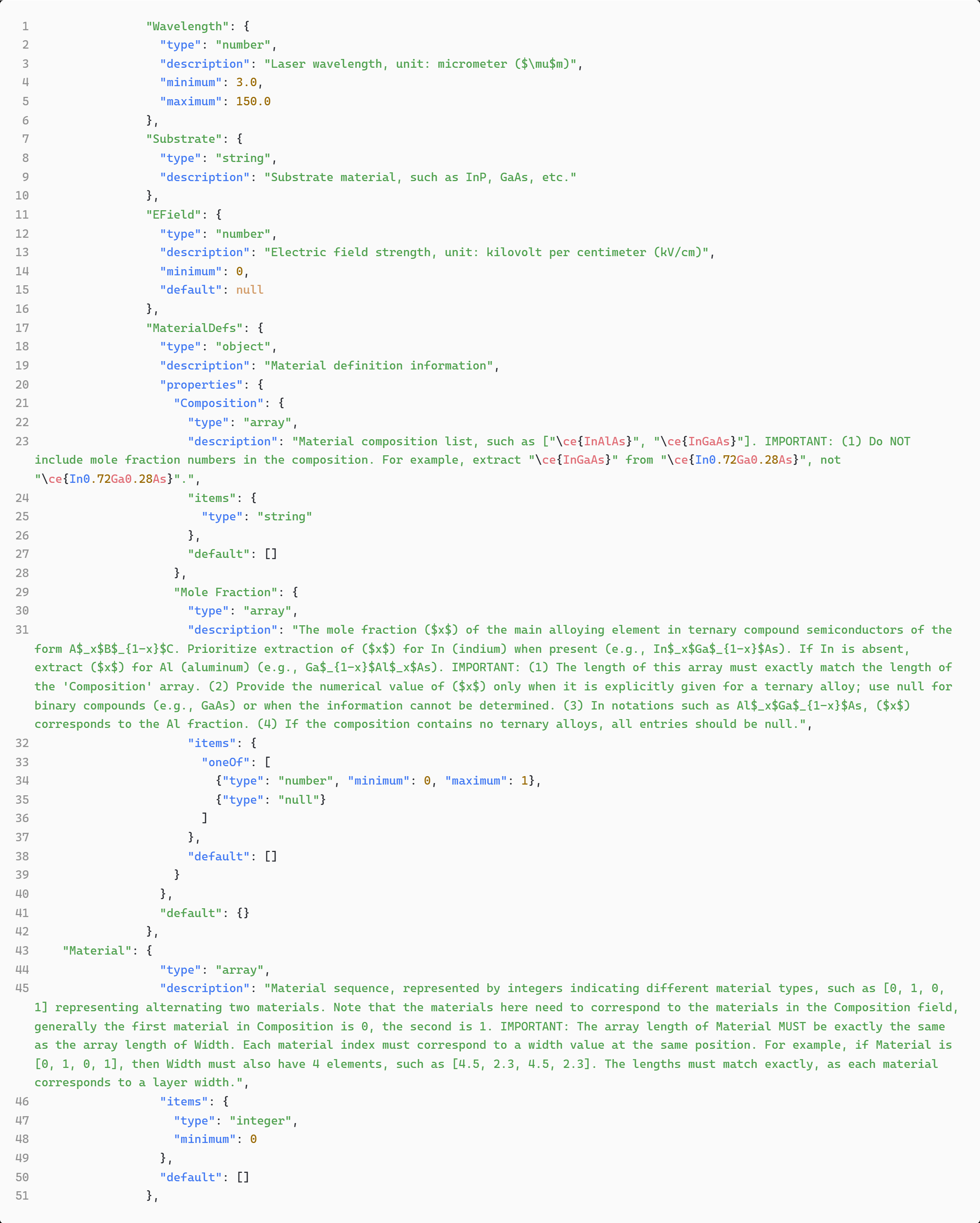}
  \caption{
  JSON Schema definition (Part II):
  Material definitions and device dimensions.}
  \label{fig:json-schema-part2}
\end{figure*}

\begin{figure*}[t]
  \centering
  \includegraphics[width=\linewidth]{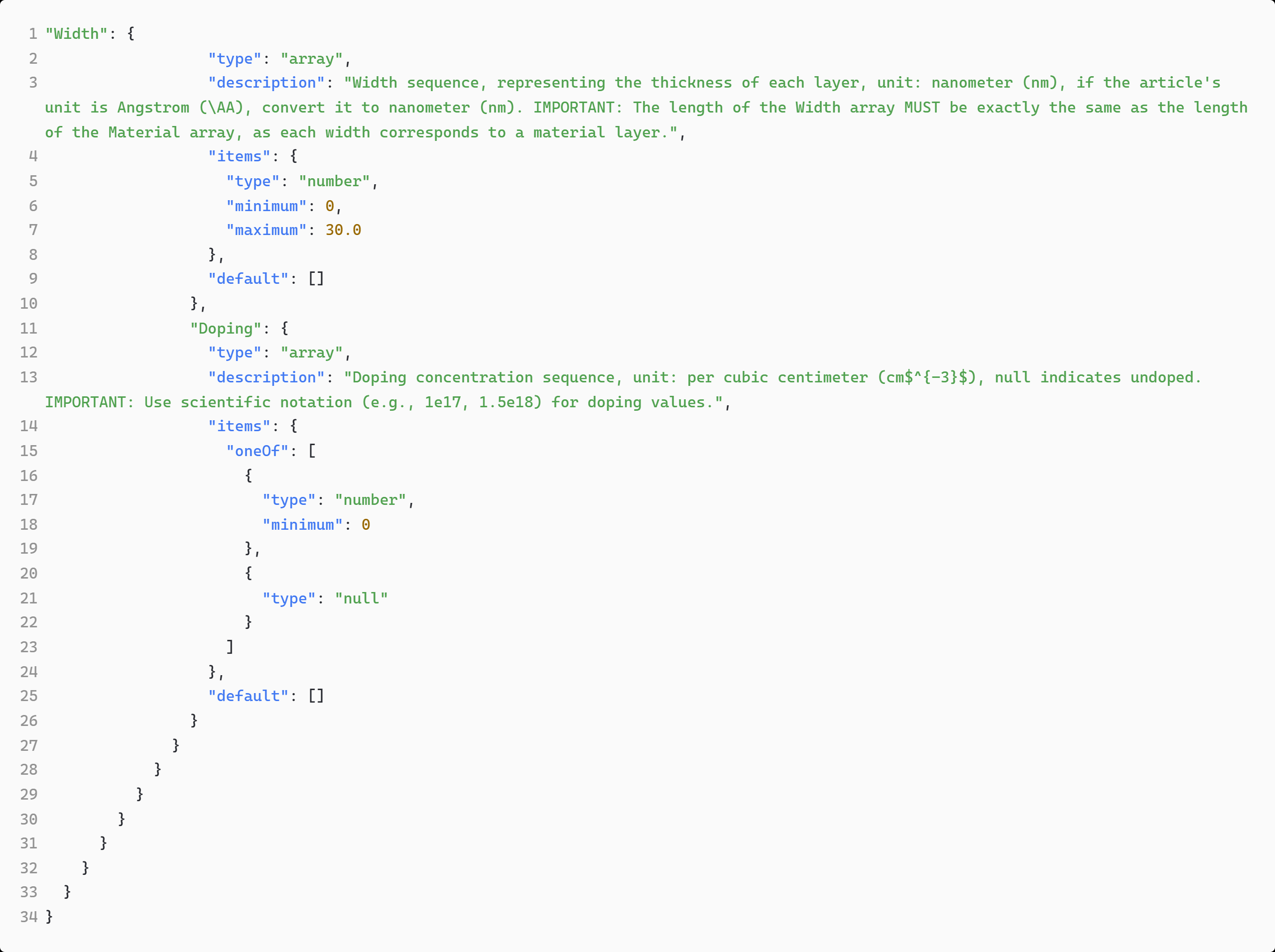}
  \caption{
  JSON Schema definition (Part III):
  Doping profiles and layer constraints.}
  \label{fig:json-schema-part3}
\end{figure*}

The formal JSON Schema definition provides precise validation rules for the extracted data.
The complete schema is presented in multiple parts due to its length,
with Figs.~\ref{fig:json-schema-part1},
\ref{fig:json-schema-part2},
and \ref{fig:json-schema-part3} showing the detailed structure definitions.

\begin{table*}[htbp]
\centering
\caption{Performance Gains Analysis: Comparative Impact of JSG-IE, Schema Formats, and Input Modalities}
\label{tab:performance_gains_analysis_detail}
\small
\begin{tabular*}{\textwidth}{@{\extracolsep\fill}l | ccc | ccc | ccc}
\toprule
\multirow{2}{*}{\textbf{Model Name}} & \multicolumn{3}{c|}{\textbf{$\Delta$ JSG-IE vs. Base}} & \multicolumn{3}{c|}{\textbf{$\Delta$ List vs. Dict}} & \multicolumn{3}{c}{\textbf{$\Delta$ Markdown vs. PDF}} \\
\cmidrule(lr){2-4} \cmidrule(lr){5-7} \cmidrule(lr){8-10}
& \textbf{P} & \textbf{R} & \textbf{$F_1$} & \textbf{P} & \textbf{R} & \textbf{$F_1$} & \textbf{P} & \textbf{R} & \textbf{$F_1$} \\
\midrule
integrated                 & +4.6 & +8.8 & +6.3 & +6.6 & +9.1 & +7.1 & +5.4 & +7.8 & +5.9 \\
kimi-k2-thinking           & +5.5 & +13.3 & +10.4 & +7.1 & +11.2 & +9.6 & +9.4 & +13.3 & +12.5 \\
gemini-3-pro-preview       & -4.7 & +3.2 & +0.6 & -5.7 & -7.6 & -5.4 & -3.9 & -3.8 & -4.5 \\
deepseek-v3-2-thinking     & +8.3 & +1.7 & +2.5 & +11.8 & +3.5 & +5.3 & +1.6 & +7.2 & +3.4 \\
deepseek-chat              & -6.3 & -7.8 & -6.9 & -2.7 & +0.7 & +1.0 & +2.4 & +0.8 & +1.2 \\
gpt-5-chat                 & -11.4 & -8.0 & -9.8 & -6.0 & -7.1 & -6.6 & +5.9 & +3.3 & +4.9 \\
deepseek-v3-2              & +0.5 & +3.2 & +2.6 & +9.0 & +10.7 & +9.6 & +8.1 & +8.2 & +7.9 \\
claude-sonnet-4-5          & +3.2 & +1.7 & +2.6 & -5.2 & -6.5 & -5.6 & +3.6 & +4.5 & +4.1 \\
qwen3-vl-235b-thinking     & +13.6 & +19.2 & +17.7 & +4.7 & +8.4 & +6.1 & +10.6 & +9.2 & +9.3 \\
qwen3-max                  & +7.1 & +9.4 & +7.2 & -9.1 & -3.1 & -7.2 & -10.0 & -7.1 & -10.3 \\
kimi-k2-0905-preview       & +9.9 & +21.4 & +18.5 & +2.9 & +3.4 & +3.2 & +4.3 & +6.1 & +5.1 \\
gpt-4o-2024-11-20          & +4.1 & +10.1 & +8.2 & -5.2 & -2.7 & -2.7 & +2.7 & +6.3 & +5.2 \\
glm4-6                     & +23.5 & +25.0 & +24.1 & +15.1 & +24.3 & +22.2 & +9.3 & +19.8 & +16.2 \\
\midrule
\textbf{Average}    & \pmb{+4.0} & \pmb{+7.7} & \pmb{+5.7} & \pmb{+1.9} & \pmb{+3.1} & \pmb{+1.8} & \pmb{+4.6} & \pmb{+6.8} & \pmb{+5.1} \\
\bottomrule
\end{tabular*}
\end{table*}

\begin{sidewaystable*}[htbp]
\centering
\caption{Performance Comparison across Different Schema Formats}
\label{tab:schema_comparison_2}
\small
\begin{tabular*}{\textheight}{@{\extracolsep\fill}l | ccc | ccc | ccc | c}
\toprule
\multirow{2}{*}{\textbf{Model Name}} & \multicolumn{3}{c|}{\textbf{Baseline}} & \multicolumn{3}{c|}{\textbf{Dict of List}} & \multicolumn{3}{c|}{\textbf{List of Dict}} & \textbf{Avg} \\
\cmidrule(lr){2-4} \cmidrule(lr){5-7} \cmidrule(lr){8-10} \cmidrule(lr){11-11}
& \textbf{P} & \textbf{R} & \textbf{$F_1$} & \textbf{P} & \textbf{R} & \textbf{$F_1$} & \textbf{P} & \textbf{R} & \textbf{$F_1$} & $F_1$\\
\midrule
\textbf{integrated} & 77.8 & 65.2 & 70.7 & 75.1 & 71.1 & 72.8 & \pmb{84.0} & 82.2 & 81.5 & 75.0 \\
\midrule
\textbf{kimi-k2-thinking} & 72.4 & 60.1 & 65.2 & 77.1 & 66.3 & 70.7 & 82.2 & \pmb{87.5} & \pmb{83.4} & \pmb{73.1} \\
\textbf{gemini-3-pro-preview} & 74.9 & \pmb{62.7} & 67.8 & 75.2 & 72.3 & \pmb{73.1} & 73.8 & 73.8 & 73.1 & \pmb{71.3} \\
\textbf{deepseek-v3-2-thinking} & \pmb{77.8} & 61.4 & \pmb{68.0} & \pmb{78.6} & 63.6 & 69.1 & 81.1 & 72.3 & 71.9 & \pmb{69.7} \\
\textbf{deepseek-chat} & 60.2 & 58.7 & 59.4 & 63.1 & 58.8 & 59.2 & 63.1 & 61.3 & 61.3 & 60.0 \\
\textbf{gpt-5-chat} & 57.8 & 56.8 & 57.2 & 57.0 & 58.0 & 56.8 & 57.4 & 60.1 & 57.5 & 57.2 \\
\textbf{deepseek-v3-2} & 72.5 & 57.7 & 63.4 & 64.7 & 62.1 & 63.1 & 70.8 & 70.6 & 69.5 & 65.3 \\
\textbf{claude-sonnet-4-5} & 62.1 & 61.7 & 61.9 & 69.5 & 68.6 & 69.0 & 63.5 & 63.9 & 63.4 & 64.8 \\
\textbf{qwen3-vl-235b-thinking} & 72.6 & 57.1 & 63.4 & 73.4 & \pmb{73.7} & 69.1 & 78.6 & 80.2 & 77.5 & 70.0 \\
\textbf{qwen3-max} & 56.1 & 55.6 & 55.8 & 66.0 & 65.2 & 65.6 & 65.9 & 69.5 & 65.8 & 62.4 \\
\textbf{kimi-k2-0905-preview} & 51.3 & 50.8 & 51.0 & 63.3 & 63.1 & 63.2 & 75.2 & 71.6 & 72.6 & 62.3 \\
\textbf{gpt-4o-2024-11-20} & 43.9 & 43.7 & 43.8 & 58.5 & 59.8 & 58.6 & 61.4 & 63.1 & 62.1 & 54.8 \\
\textbf{glm4-6} & 51.2 & 50.0 & 50.6 & 57.4 & 56.5 & 56.9 & 78.6 & 74.9 & 75.8 & 61.1 \\
\midrule
\textbf{Average} & \textbf{68.0} & \textbf{63.9} & \textbf{64.7} & \textbf{70.7} & \textbf{64.5} & \textbf{65.2} & \textbf{72.0} & \textbf{71.6} & \textbf{70.4} & \textbf{66.8} \\
\bottomrule
\end{tabular*}
\end{sidewaystable*}

\end{document}